\documentclass[twocolumn,showpacs,preprintnumbers,amsmath,amssymb]{revtex4}

\usepackage{graphicx}
\usepackage{dcolumn}
\usepackage{bm}
\usepackage{subfigure}

\newcommand{\pro}[2]{\langle{#1}|{#2}\rangle}
\newcommand{\bra}[1]{\langle{#1}|}
\newcommand{\ket}[1]{|{#1}\rangle}

\newcommand{\Tr}{{\rm Tr}\hspace{0.07cm}}

\begin{document}

\title{Robust Adaptive measurement for qubit state preparation }

\author{Saki Tanaka}%
\email[E-mail address: ]{saki-tanaka@a6.keio.jp}
\author{Naoki Yamamoto}
\email[E-mail address: ]{yamamoto@appi.keio.ac.jp}
\affiliation{Department of Applied Physics and Physico-Informatics, 
Keio University, Yokohama 223-8522, Japan}
\date{\today}%

\begin{abstract}

This paper reconsiders the method of adaptive measurement for 
qubit state preparation developed by Jacobs and shows an 
alternative scheme that works even under unknown unitary evolution 
of the state. 
The key idea is that the measurement is adaptively changed so that 
one of the eigenstates of the measured observable is always set 
between the current and the target states at while that eigenstate 
converges to the target. 
The most significant feature of this scheme is that the measurement 
strength can be taken constant unlike Jacobs' one, which eventually 
provides fine robustness property of the controlled state against 
the uncertainty of the unitary evolution. 

\end{abstract}


\maketitle


\section{Introduction}

Repeated measurement of an observable that is appropriately changed 
according to the pre-measurement outcomes, i.e., the {\it adaptive 
measurement}, has great potential for various purposes in quantum 
information sciences. 
The first demonstration has come out in the application to quantum 
phase estimation 
\cite{WisemanCollections,ArmenPRL2002,TsangPRA2009,WheatleyPRL2010,
YonezawaScience2012}. 
Another application of adaptive measurement is for state 
preparation \cite{JacobsNJP2010,AshhabPRA2010,WisemanNature2011,
CiccarelloPRL2008,YuasaNJP2009}. 
A striking feature of this scheme is that a desired time evolution 
of the state is brought only by measurement back-action, and there 
is no need to introduce any external force for controlling the 
state.

Let us especially focus on the method developed by Jacobs 
\cite{JacobsNJP2010}. 
This employs the schematic of a continuous-time measurement; 
in this case the probabilistic change of a qubit state $\hat\rho$ 
is described by the following {\it stochastic master equation} 
(SME) \cite{Handel,QM}:
\begin{align}
\label{eq:timeevadaptive}
   d \hat{\rho}_t
     =& - k_t \bigl[\hat{\sigma}_t, 
               [ \hat{\sigma}_t, \hat{\rho}_t ]\bigr]dt  
                  \notag \\
     & + \sqrt{2 k_t}
       \big( \hat{\sigma}_t \hat{\rho}_t + \hat{\rho}_t \hat{\sigma}_t 
               - 2 \Tr(\hat{\sigma}_t \hat\rho_t)\hat{\rho}_t \big) dW_t, 
\end{align}
where $dW_t$ is the standard Wiener process satisfying the Ito 
rule $dW_t^2=dt$. 
In the above equation, $\hat\sigma_t$ and $k_t$ represent the 
measured observable and the measurement strength, respectively. 
Adaptive measurement means that we can change $\hat\sigma_t$ 
and $k_t$ continuously in time, as functions of the state 
$\hat\rho_t$, so that $\hat\rho_t$ will converge to a target state. 
In Jacobs' scheme, the state is assumed to be pure (this actually 
holds if the initial state is pure) and is thus of the form 
\begin{equation}
\label{Bloch representation}
   \hat\rho_t=\ket{\psi_t}\bra{\psi_t},~~~
   \ket{\psi_t} = \cos(\delta_t /2 ) \ket{0} + \sin(\delta_t /2) \ket{1}, 
\end{equation}
where $\ket{0}=(1,0)^\top$ and $\ket{1}=(0,1)^\top$ are the 
target state and the initial state, respectively. 
Then the observable and the measurement strength are updated 
according to the following laws:
\begin{align}
\label{eq:Jacobsaxis}
    \hat{\sigma}_t^{\rm J} 
       = \hat{\sigma}_x \cos(\delta_t) - \hat{\sigma}_z \sin(\delta_t),~~~
    k_t^{\rm J}=\kappa \delta_t^2, 
\end{align}
where $\kappa$ is a positive constant and $\hat{\sigma}_{i}$ is the 
Pauli matrix. 
(The index ``${\rm J}$'' indicates that it is the scheme proposed by Jacobs.) 
This means that the measured observable $\hat\sigma_t^{\rm J}$ 
is changed so that its eigenstates are always perpendicular to 
the current state $\ket{\psi_t}$ in the Bloch sphere representation, 
as shown in Fig.~1~(a). 
The measurement strength $k_t^{\rm J}$ is also adaptively changed, 
and it decreases proportionally to the distance between the current 
and target states; 
note that the measurement becomes very weak when the state 
approaches to the target. 
In fact, with this adaptive measurement law \eqref{eq:Jacobsaxis}, 
the time evolution of $\delta_t\in[0,\pi]$ is given by 
\begin{align}
\label{eq:Jacobsdelta}
   d\delta_t = \sqrt{8 k_t^{\rm J}} dW _t
             = \sqrt{8\kappa} \delta_t dW_t, 
\end{align}
and it was numerically shown in \cite{JacobsNJP2010} that $\delta_t$ 
converges to zero as $t\rightarrow \infty$ almost surely.

\begin{figure}[b]
 \centering 
 \includegraphics[width=8.5cm]{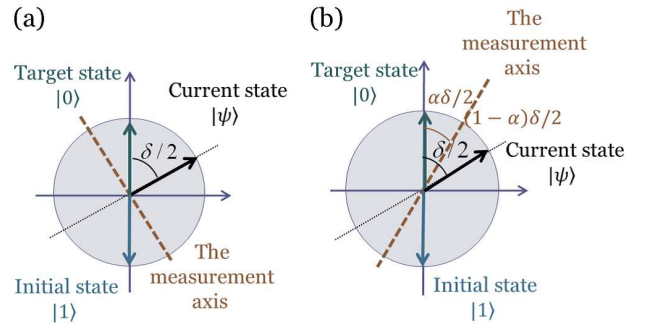}
 \caption{
 (Color online) 
 The measurement axes of Jacobs' adaptive measurement scheme (a) 
 and the presented scheme (b). 
 } 
 \label{fig:axises}
\end{figure}

Here we come up with the question about how much the above 
adaptive measurement scheme is robust against certain disturbance 
acting on the system. 
Actually such robustness against a specific decohering effect 
was evaluated in \cite{JacobsNJP2010}. 
Another specific but important disturbance is an {\it unknown} 
unitary evolution of the qubit state, in which case the driving 
term $-i[\hat H, \hat\rho_t]dt$ is added to the right hand side 
of Eq.~\eqref{eq:timeevadaptive}. 
For example, if we take a two-level atom continuously observed 
using the Faraday rotation technique \cite{Stockton} to realize 
the qubit system subjected to Eq.~\eqref{eq:timeevadaptive}, such 
a disturbing Hamiltonian may appear and take the form 
$\hat H=\Delta\hat\sigma_y$ with $\Delta$ unknown detuning between 
the atomic transition frequency and the laser frequency. 
Note that in \cite{JacobsNJP2010} this kind of unknown disturbance 
was not discussed.

In this paper, we reconsider the same control problem discussed 
above, yet with additional care about the influence of an unknown 
unitary evolution, and then propose a new adaptive measurement 
scheme that has clear robustness property against that disturbance. 
A novel difference between our scheme and Jacobs' one is that 
we take the measurement axis between the currenet and the target 
states in the sense shown in Fig.~\ref{fig:axises}~(b), rather than 
the orthogonal one;
then that measurement axis converges to the target, so that the 
state may be probabilistically moved toward the target as well. 
The main feature of this scheme is that we can keep the measurement 
strength constant during all the time-evolution, while in Jacobs' 
case it must be weakened when the state approaches to the target. 
This mechanism brings the following two merits: 
(i) The first is that the experimental implementation becomes 
simpler; 
in fact we need to adaptively change only the measurement axis. 
(ii) Secondly, because the measurement strength is not weakened, 
the system preserves capability moving the state via measurement 
back action even near the target. 
The latter is more important in the present context, because it 
is then expected that the adaptive measurement can deal with the 
unknown force along all the time evolution, while in Jacobs' case 
the unknown force has to become dominant and thus cannot be 
suppressed when the state approaches to the target. 
This fact will be actually demonstrated in numerical simulations.


\section{The adaptive measurement scheme}
\label{zenoeffectcontrolling}

In general, if we continuously measure an observable in a QND manner, 
the state moves probabilistically toward one of the eigenstates of 
that observable \cite{Handel,QM}. 
Hence, if this measured observable is changed adaptively so that 
the eigenstate attracting the current state approaches to a 
desired target state, it is expected that the state will finally 
be stabilized at that target state. 
Based on this idea, we take the following observable as a 
measured observable in Eq.~\eqref{eq:timeevadaptive}: 
\begin{align}
\label{eq:dragaxis}
     \hat{\sigma}_t^{\rm R}
         =\hat{\sigma}_x \sin(\alpha \delta_t) 
                 + \hat{\sigma}_z \cos(\alpha \delta_t), 
\end{align}
where $\alpha\in[0,1]$ is the tuning parameter whose meaning is 
explained below. 
The index ``${\rm R}$'' indicates that it is the ``Robust" adaptive 
measurement scheme to discern it from Jacobs' one. 
The eigenstates of the observable $\hat{\sigma}_t^{\rm R}$ are 
given by 
\begin{align*}
    \ket{+_{\rm{Z}}} 
      =  \begin{pmatrix} 
            \cos (\alpha \delta_t/2 ) \\ 
            \sin (\alpha \delta_t /2)  
         \end{pmatrix}, ~~
    \ket{-_{\rm{Z}}} 
      =  \begin{pmatrix} 
            \sin (\alpha \delta_t/2 ) \\ 
            -\cos (\alpha \delta_t /2)  
         \end{pmatrix}, 
\end{align*}
which satisfy 
$\hat{\sigma}_t^{\rm R} \ket{+_{\rm Z}} = \ket{+_{\rm{Z}}},~ 
 \hat{\sigma}_t^{\rm R} \ket{-_{\rm Z}} = -\ket{-_{\rm Z}}$. 
In the Bloch sphere representation, $\ket{+_{\rm{Z}}}$ divides the 
angle between the target state $\ket{0}$ and the current state 
$\ket{\psi_t} = (\cos(\delta_t/2),~\sin(\delta_t/2) )^\top$ 
into $\alpha :(1-\alpha )$; see Fig.~\ref{fig:axises}~(b). 
The transition probabilities of the state jumping to these 
eigenstates are 
$p_+=|\pro{ +_{\rm Z} }{\psi_t}|^2 = (1 + \cos(\beta \delta_t)/2$ 
and 
$p_-=|\pro{ -_{\rm Z} }{\psi_t}|^2 = (1 - \cos(\beta \delta_t))/2$, 
where $\beta := 1-\alpha$, and they satisfy $p_+\geq p_-$ if 
$\cos(\beta \delta_t) \geq 0$. 
This means that the state tends to move towards the direction 
of $\ket{+_{\rm Z}}$ with probability $p_+$.
At the same time, since in this case $\delta_t$ decreases, the 
eigenstate $\ket{+_{\rm{Z}}}$ moves towards the target $\ket{0}$. 
In particular, when the state reaches the target $\ket{0}$, 
or equivalently $\delta_t\rightarrow 0$, the eigenstate 
$\ket{+_{\rm Z}}$ becomes identical to the target and the 
transition probability $p_+$ takes the value $1$; 
that is, the state is stabilized at the target as expected.

Next, the measurement strength is simply set to a constant value 
$k_t^{\rm R} = k^{\rm R}$, unlike the case of Jacobs' scheme. 
The reason will be explained in Remark (i) given in the end of 
this section, but we here point out that this setting has a clear 
merit from a practical viewpoint. 
In fact, changing the measurement strength in addition to changing 
the measured observable definitely costs more expensive compared 
to the case where only the latter is required. 
In this sense, our scheme is suited to experimental implementation.

Regarding the disturbing Hamiltonian, as discussed in Sec.~I, 
we choose $\hat{H}= \Delta \hat{\sigma}_y$, where $\Delta$ is an 
unknown constant. 
Note that with this Hamiltonian the state rotates around the $y$ 
axis in the Bloch sphere.

Consequently, the dynamical evolution of the state under the 
adaptive measurement setup and the disturbing Hamiltonian 
introduced above is given by 
\begin{align}
\label{eq:timeevadaptiveforce}
    d \hat{\rho}_t 
       = & - i \Delta [\hat{\sigma}_y, \hat{\rho}_t]dt 
           - k^{\rm R} [\hat{\sigma}_t^{\rm R}, 
                   [ \hat{\sigma}_t^{\rm R}, \hat{\rho}_t ] ] dt 
\notag \\
     \mbox{}
       & + \sqrt{2k^{\rm R}} 
         \big( \hat{\sigma}_t^{\rm R} \hat{\rho}_t 
           + \hat{\rho}_t \hat{\sigma}_t^{\rm R} 
            - 2 \Tr(\hat{\sigma}_t^{\rm R} \hat\rho_t)\hat\rho_t \big) dW_t. 
\end{align}
The initial state is now on the $x$-$z$ plane, hence 
we have the dynamics of $\delta_t$ as follows: 
\begin{align}
\label{eq:deltadtimeev}
    d\delta_t = 2 \Delta dt -2k^{\rm{R}} \sin(2 \beta \delta_t ) dt 
               + \sqrt{8k^{\rm{R}}} \sin( \beta \delta_t) dW_t. 
\end{align}
The linear approximated equation, which is valid when 
$\delta_t \approx 0$, is 
\begin{align}
\label{eq:deltadtimeev1ap}
     d\delta _t = 2 \Delta dt - 4k^{\rm{R}} \beta \delta_t dt 
                    + \sqrt{8k^{\rm{R}}} \beta \delta_t dW_t. 
\end{align}
On the other hand, the dynamics of the same $\delta_t$ but with 
Jacobs' scheme is given by 
\begin{equation} 
\label{JacobsdeltaDelta}
    d\delta _t 
       = 2\Delta dt + \sqrt{8\kappa} \delta_t dW_t. 
\end{equation}

A striking difference of the above two dynamics 
\eqref{eq:deltadtimeev1ap} and \eqref{JacobsdeltaDelta} is that 
the former contains an additional drift term 
$-4k^{\rm R}\beta \delta_t dt$ while in the latter equation there 
is no such state-dependent term. 
Note that this additional drift term apparently works for driving 
$\delta_t$ toward zero. 
This fine property of our scheme is brought from the mechanism 
that, around $\delta_t=0$, the target state itself is continuously 
measured and thus the state is attracted to the target with very 
high probability. 
In contrast, as mentioned before, in Jacobs' case the measurement 
has to be weakened and finally turned off when the state reaches 
the target, thus there is no such attracting effect changing 
$\delta_t$ to zero. 
We further expect that the additional drift term in 
Eq.~\eqref{eq:deltadtimeev1ap} implies no more than the 
enhancement of stability of the dynamics of $\delta_t$, 
which consequently makes the system robust against the disturbing 
noise. 
Note that this observation makes sense only when the state 
is around the target. 
Therefore, in the later sections we will examine some numerical 
simulations to actually verify the above-mentioned driving effect 
and resulting robustness property.

Before closing this section, we provide two remarks.

{\it Remark} (i):
Jacobs' scheme requires the adaptive tuning of the measurement 
strength (i.e., $k_t^{\rm J}=\kappa \delta_t^2$ in 
Eq.~\eqref{eq:Jacobsaxis}) for the dynamics of $\delta_t$ to 
have the state-dependent diffusion term. 
Such state-dependence is indeed necessary to generate dynamical 
stability of $\delta_t$. 
Hence, it should be maintained that, with our measurement 
scheme, the diffusion term of Eq.~\eqref{eq:deltadtimeev} 
depends on the state even with the fixed measurement strength 
(i.e., $k_t^{\rm R} = k^{\rm R}$).

{\it Remark} (ii): 
The time evolution of the fidelity between the current state 
$\hat \rho_t$ and the target state $\ket{0}$ is given by 
\begin{align}
   d\bra{0}\hat \rho_t\ket{0} 
     = \Big[ -\Delta \sin\delta_t + 
         k^{\rm R} \Big( \cos(2\alpha\delta_t-\delta_t) 
               - \cos\delta_t \Big)  \Big]dt 
\notag \\
     - \sqrt{2 k^{\rm R}} \Big[ 
           \frac{3+ \cos\delta_t }{2} \cos(\alpha\delta_t -\delta_t) 
            + \cos(\alpha \delta_t ) \Big] dW_t. 
\label{eq_cost}
\end{align}
Hence, the optimum value of $\alpha$ that maximizes the deterministic 
change per unit time of the fidelity is given by $\alpha=1/2$. 
This value is actually taken in the simulations shown later.


\section{State convergence}

\begin{figure}[htbp!]
 \centering
 \includegraphics[width=.48\textwidth]{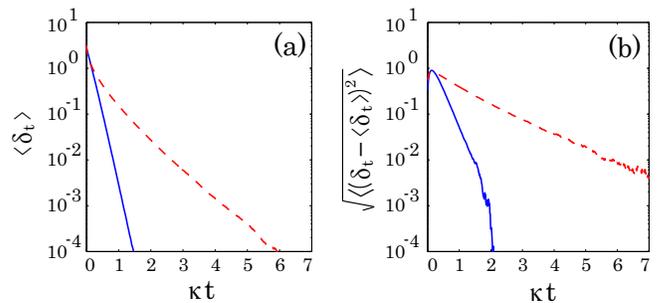}
 \caption{
 (Color online) 
 Time evolutions of (a) the mean and (b) the standard deviation of 
 $\delta_t$ when $\Delta=0$. 
 The solid blue line corresponds to our adaptive measurement scheme 
 while the dashed red line does Jacobs' scheme. }
 \label{Delta0}
\end{figure}

\begin{figure}[htbp!]
 \centering
 \includegraphics[width=.48\textwidth]{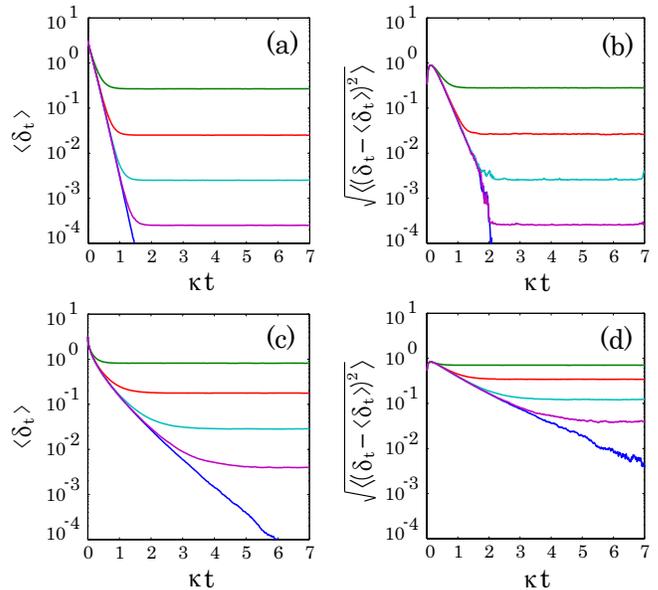}
 \caption{
 (Color online) 
 Time evolutions of (a) the mean and (b) the standard deviation 
 of $\delta_t$ subjected to our scheme, for several values of $\Delta$. 
 The green, red, water blue, purple, and blue lines correspond to 
 $\Delta=1, 1/10, 1/10^2, 1/10^3$, and $\Delta=0$, respectively. }
 \label{DeltaChanged}
\end{figure}

In this section, to verify that our adaptive measurement scheme 
actually works for driving the state towards the target, we examine 
some numerical simulations, under the assumption that $\Delta$ is 
{\it known}; 
this is a crucial assumption because we are then able to update 
$\delta_t$ and as a result the adaptive measurement law 
\eqref{eq:dragaxis} (or Eq.~\eqref{eq:Jacobsaxis}) exactly 
by recursively solving Eq.~\eqref{eq:deltadtimeev} 
(or Eq.~\eqref{JacobsdeltaDelta} for Jacobs' case). 
For reference, we show the trajectories in Jacobs' case as well; 
in particular, we set the same diffusion coefficients in 
Eqs.~\eqref{eq:deltadtimeev1ap} and \eqref{JacobsdeltaDelta}; 
i.e., $\sqrt{8 k^{\rm R}}\beta=\sqrt{8 \kappa}$, 
which leads to $k^{\rm R}\beta^2=\kappa$. 
Hence let us here take the parameters as $\kappa=1$, $k^{\rm R}=4$, 
and $\alpha=1-\beta=1/2$ (see Remark (ii) in Sec.~II). 
The disturbance strength $\Delta$ takes several values. 
The initial condition is $\delta_0=\pi$, as defined below 
Eq.~\eqref{Bloch representation}.

Figure \ref{Delta0} shows the time evolutions of the mean and the 
standard deviation of $\delta_t$, when $\Delta=0$. 
The plots are obtained by averaging $10^6$ sample paths. 
The solid blue and the dashed red lines correspond to the cases of 
our adaptive measurement scheme and Jacobs' one, respectively. 
We find from the figures that, in both cases, the state certainly 
converges to the target with almost probability one. 
Note that the plots do not imply that our scheme offers faster 
and stable convergence of the state compared to Jacobs' case; 
this is because the measurement strength of Jacobs' scheme has 
to be weakened, implying slower change of the state around the 
target.

Next, Fig.~\ref{DeltaChanged} shows the means and the standard 
deviations of $\delta_t$, with several values of $\Delta$. 
As expected, the disturbance prevents the state from converging 
to the target. 
Notably, in the long time limit, the error is approximately 
proportional to $\Delta$. 
Similar plots are obtained in Jacobs' case as well.

{\it Remark}: 
Although we mentioned above that the figures do not imply the 
superiority of our scheme over Jacobs' one, a trivial fair 
comparison can be performed as follows. 
In fact, if we take a constant measurement strength 
$k_t^{\rm J} = k^{\rm J}$ in Jacobs' scheme, the dynamics is given 
by $d\delta_t = \sqrt{8 k^{\rm J}} dW _t$; 
clearly, then, $\delta_t$ does not converge to zero even under
the boundary condition. 
That is, if we run the two schemes with the same constant 
measurement strength, clearly our scheme offers better performance 
over Jacobs' one.


\section{Robustness of the adaptive Zeno measurement}

Here we study the case where the disturbance magnitude $\Delta$ is 
{\it unknown}. 
To make the situation clear, let us again consider the general 
SME \eqref{eq:timeevadaptive} that is additionally driven by 
an unknown Hamiltonian $\hat H$: 
\begin{align}
\label{true SME}
   d \hat{\rho}_t
    =& -i [\hat H, \hat\rho_t ]dt 
     - k_t \big[\hat{\sigma}_t, [ \hat{\sigma}_t, \hat{\rho}_t ]\big]dt  
\notag \\
     & + \sqrt{2 k_t}
       \big[ \hat{\sigma}_t \hat{\rho}_t + \hat{\rho}_t \hat{\sigma}_t 
          - 2 \Tr(\hat{\sigma}_t \hat\rho_t)\hat{\rho}_t \big] dW_t. 
\end{align}
This {\it true state} $\hat\rho_t$ cannot be precisely updated, 
due to the uncertainty of $\hat H$. 
Therefore, we need to devise an updating law of a {\it nominal state}, 
say $\hat\rho_t'$, only using the measurement result $y_t$ that 
is subjected to the output equation
\begin{equation}
\label{output}
   dy_t = \Tr(\hat{\sigma}_t \hat{\rho}_t)dt + dW_t. 
\end{equation}
Note this is driven by the same $dW_t$ as that in Eq. \eqref{true SME}; 
$dW_t$ is called the {\it innovation} in the framework of quantum 
filtering theory \cite{Belavkin,Bouten}. 
To update $\hat\rho_t'$, we particularly use Eqs.~\eqref{true SME} 
and \eqref{output} with $\hat H$ replaced by a known nominal 
Hamiltonian $\hat{H}'$ as follows: 
\begin{align}
\label{nominal SME}
   d \hat{\rho}'_t
    =& -i [\hat H', \hat\rho_t' ]dt 
     - k_t \big[\hat{\sigma}_t, [ \hat{\sigma}_t, \hat{\rho}'_t ]\big]dt  
\notag \\
     & + \sqrt{2 k_t}
       \big[ \hat{\sigma}_t \hat{\rho}'_t + \hat{\rho}'_t \hat{\sigma}_t 
               - 2 \Tr(\hat{\sigma}_t \hat\rho'_t)\hat{\rho}'_t \big]
\notag \\
     & \hspace{3em} \times [dy_t - \Tr(\hat{\sigma}_t \hat{\rho}'_t)dt]. 
\end{align}
Note again that $y_t$ is the measurement result and is thus known. 
Hence we can recursively calculate the nominal state $\hat\rho_t'$, 
although it should differ from the true state $\hat\rho_t$. 
In the adaptive measurement setup, therefore, the observable 
$\hat\sigma_t$ and the strength $k_t$ are changed in time as 
functions of $\hat\rho_t'$.

For the specific problem under consideration, let 
$\hat{H} = \Delta \hat{\sigma}_{\rm y}$ and 
$\hat{H}' = \Delta' \hat{\sigma}_{\rm y}$ be the true and nominal 
Hamiltonians, respectively; 
$\Delta$ is an unknown constant while $\Delta'$ is a known nominal 
constant. 
Also we define $\delta_t$ and $\delta'_t$, corresponding to the 
true and the nominal states. 
In our measurement scheme, these variables are driven by the 
following equations: 
\begin{eqnarray}
& & \hspace*{-3em}
\label{true Zeno update rule}
    d\delta_t 
       = \big[ 2 \Delta
           - 2 k^{\rm Z} \sin(2\delta_t-2\alpha \delta_t') \big]dt
\nonumber \\ & & \hspace*{1em}
    \mbox{}
      + \sqrt{8k^{\rm Z}} \sin(\delta_t - \alpha \delta_t') dW_t, 
\\ & & \hspace*{-3em}
\label{nominal Zeno update rule}
      d \delta_t' 
       = \big[2 \Delta' 
           -2 k^{\rm Z}\sin(2 \beta \delta_t')\big]dt 
\nonumber \\ & & \hspace*{1em}
    \mbox{}
        + \sqrt{8k^{\rm Z}} \sin( \beta \delta_t') dW_t^{\rm Z}, 
\end{eqnarray}
where
\begin{equation*}
   dW_t^{\rm Z} = 
      \big[ \cos(\beta\delta_t') - \cos(\delta_t'-\alpha \delta_t) \big]dt
         + dW_t. 
\end{equation*}
Note that $\beta=1-\alpha$. 
The above two equations \eqref{true Zeno update rule} and 
\eqref{nominal Zeno update rule} take the same form when 
$\Delta= \Delta'$ and $\delta_t=\delta_t'$. 
Jacobs' scheme, on the other hand, leads to the following 
equations to update $\delta_t$ and $\delta_t'$: 
\begin{eqnarray}
& & \hspace*{-3em}
\label{true Jacobs update rule}
    d \delta_t
     = \big[2 \Delta
      + 2 \kappa \delta_t'\mbox{}^2 \sin(2 \delta_t -2 \delta_t')\big]dt 
\nonumber \\ & & \hspace*{1em}
    \mbox{}
     + \sqrt{8\kappa} \delta_t'\cos( \delta_t - \delta_t')dW_t, 
\\ & & \hspace*{-3em}
\label{nominal Jacobs update rule}
    d \delta_t' 
       = 2 \Delta' dt 
           + \sqrt{8\kappa}  \delta_t' dW_t^{\rm J},
\end{eqnarray}
where
\begin{equation*}
    dW_t^{\rm J} =  \sin( \delta_t' -\delta_t ) dt + dW_t. 
\end{equation*}
Again, the above two equations \eqref{true Jacobs update rule} and 
\eqref{nominal Jacobs update rule} take the same form when 
$\Delta=\Delta'$ and $\delta_t=\delta_t'$.

\begin{figure}[t]
 \centering
 \includegraphics[width=.47\textwidth]{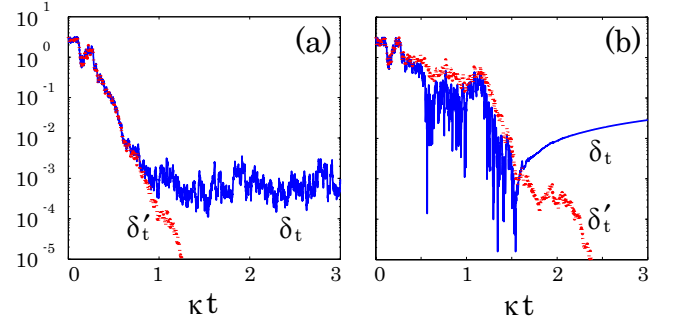}
 \caption{
 (Color online) 
 Sample paths of the true variable $\delta_t$ (solid blue lines) 
 and the nominal one $\delta_t'$ (dashed red lines) for 
 (a) our adaptive measurement scheme and (b) Jacobs' scheme, where 
 $\Delta=1/10^2$ and $\Delta'=0$.  
 }
 \label{fig:Robustness}
\end{figure}

In Fig.~\ref{fig:Robustness} we show sample paths of $\delta_t$ 
(solid blue line) and $\delta_t'$ (dashed red line) for both of 
the adaptive measurement schemes, with the same parameters taken 
in Sec.~III, fixed uncertainty $\Delta = 1/10^2$, and a typical 
nominal value $\Delta' = 0$. 
We find from the figures that, with our adaptive measurement 
scheme, the true variable $\delta_t$ is stabilized around the target 
$\delta_t=0$ though it still has a certain error corresponding to 
the constant external force, while in Jacobs' scheme $\delta_t$ 
finally follows a deterministic time-evolution and goes away 
from zero; in every numerical simulation a similar behavior is 
observed.

We here address the mechanism that brings about the above drastic 
difference. 
First, since the nominal state is now governed by the dynamics 
without uncertainty, $\delta_t'$ converges to zero, as seen in 
Sec.~III. 
When $\delta_t'\rightarrow 0$, the measured observable becomes 
$\hat\sigma_t=\hat\sigma_z$ for both the adaptive measurement 
schemes, but the measurement strength becomes $k_t^{\rm J}=0$ for 
Jacobs' case while $k_t^{R}=k^{\rm R}$ in our case; 
that is, Jacobs' scheme stops to measure the system once 
$\delta_t'=0$ is reached, and consequently the true variable 
$\delta_t$ turns out to follow the deterministic dynamics 
$d\delta_t/dt=2\Delta$. 
In contrast, our scheme keeps measuring $\sigma_z$, even after 
$\delta_t'$ takes zero, implying that the true state can be 
stabilized around the target.

\begin{figure}[tbp!]
 \centering
 \includegraphics[width=.47\textwidth]{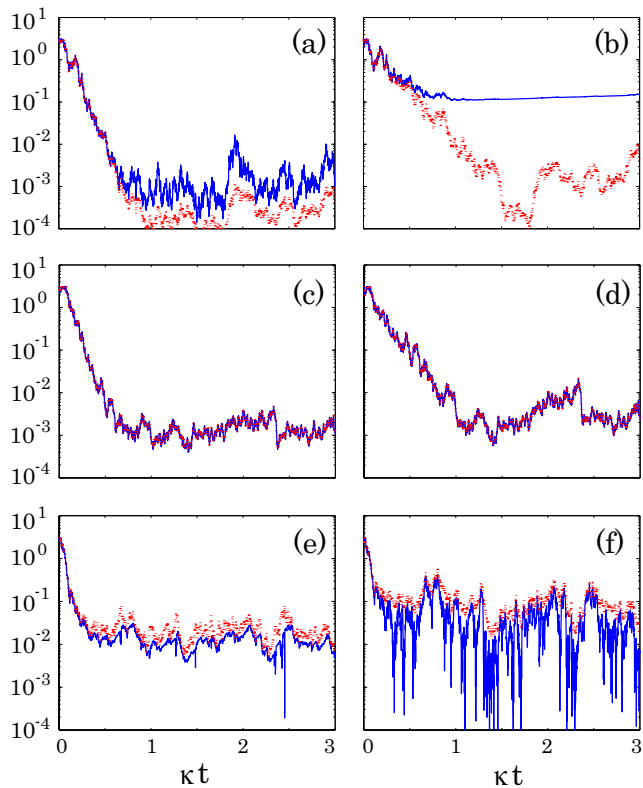}
 \caption{
 (Color online) 
 Sample paths of the true variable $\delta_t$ (solid blue lines) 
 and the nominal one $\delta_t'$ (dashed red lines) for (a, c, e) 
 the presented adaptive measurement scheme and (b, d, f) Jacobs' 
 scheme. 
 For all cases, the true external force is set to $\Delta=1/10^2$, 
 while several nominal values of $\Delta'$ are examined: 
 (a, b) represent the case when $\Delta'=1/10^3$, 
 (c, d) for the case $\Delta'=1/10^2$, 
 and (e, f) for the case $\Delta'=1/10$. 
 }
 \label{robustnessdelta}
\end{figure}

The above statement holds only for the case of $\Delta'=0$; 
hence we take some non-zero values of $\Delta'$ and show in 
Fig.~\ref{robustnessdelta} the trajectories of $\delta_t$ 
and $\delta_t'$. 
If $\Delta'$ is not zero but smaller than the true value $\Delta$ 
(Figs.~\ref{robustnessdelta}~(a, b)), we observe similar 
trajectories to the case of $\Delta'=0$; 
that is, in Jacobs' case the true variable again goes away 
from zero deterministically, due to the same reason stated 
above. 
In Figs.~\ref{robustnessdelta}~(c, d) where $\Delta=\Delta'$, 
the nominal variable $\delta_t'$ exactly tracks $\delta_t$, 
which is the property the nominal updating laws 
\eqref{nominal Zeno update rule} and 
\eqref{nominal Jacobs update rule} should have. 
Lastly, if $\Delta'$ is bigger than $\Delta$ 
(Figs.~\ref{robustnessdelta}~(e, f)), or in other words if 
we take a conservative approach for estimating $\delta_t$, 
Jacobs' scheme can stabilize the state around the target 
although it has big fluctuation, while our measurement 
scheme moves the state more close towards zero with much 
smaller estimation error.

\begin{figure}[tbp!]
 \centering
 \includegraphics[width=.47\textwidth]{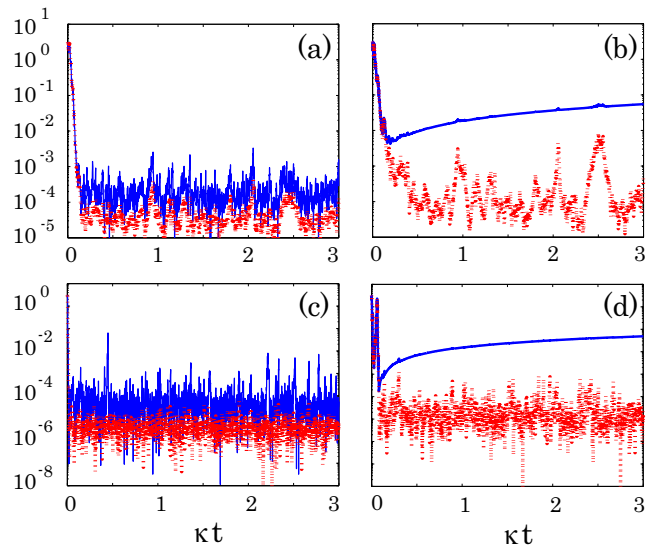}
 \caption{
 (Color online) 
 Sample paths of the true variable $\delta_t$ (solid blue lines) 
 and the nominal one $\delta_t'$ (dashed red lines) for (a, c) 
 the presented adaptive measurement scheme and (b, d) Jacobs' 
 scheme. 
 For all cases, we set $\Delta=1/10^2$ and $\Delta'=1/10^3$. 
 The measurement strength is taken as $\kappa=5$ for the cases 
 (a, b) while $\kappa=50$ for (c, d). 
 }
 \label{robustness kappa}
\end{figure}

We further expect that the issue found in Jacobs' case may be 
avoided by taking a large value of $\kappa$; 
actually, the major drawback of Jacobs' scheme is that the 
measurement strength has to be weakened around the target, which 
brings less stability of the state. 
Hence let us try some larger values of $\kappa$ to see if a similar 
issue can still happen. 
Figures~\ref{robustness kappa}~(b, d) demonstrate the nominal 
and true valus of $\delta_t$ in Jacobs' case, with the parameters 
$\Delta=1/10^2$ and $\Delta'=1/10^3$; 
in particular, here a relatively large values of $\kappa$ are taken, 
i.e., $\kappa=5$ for the case (b) and $\kappa=50$ for the case (d). 
Clearly, the stability are enhanced compared 
to the previous cases shown in Fig.~\ref{robustnessdelta} 
in the sense that the convergence speed becomes faster. 
However, the issue is not resolved; 
that is, in both cases the nominal state fails to track the 
true one once it approaches to the target, and then the true 
one obeys the deterministic unitary time evolution and flows 
away from the target. 
On the other hand, as shown in Figs.~\ref{robustness kappa}~(a, c), 
our measurement scheme does not bring such flow of the state, 
because of the intrinsic stability given to both $\delta_t$ and 
$\delta_t'$. 
Note that, when $\Delta'>\Delta$, the issue observed in 
Fig.~\ref{robustnessdelta}~(f) still happens, while in our case 
the fluctuation is well suppressed.

Summarizing, our adaptive measurement scheme offers better 
estimation of $\delta_t$, compared to Jacobs' case, in the 
sense that it is fairly robust against the uncertainty of the 
additional unitary time evolution. 
It should be maintained again that this nice stability property 
is brought from the mechanism of continuous monitoring of the 
system with constant measurement strength.


\section{Concluding remarks}

The main features of the presented adaptive measurement scheme are 
twofold; 
the first is that the measured observable contains an eigenstate 
that eventually converges to the target state, and the other is 
that the measurement strength can be taken constant even when 
the state is close to the target, which is the crucial property 
bringing the robustness property. 
As long as the above two conditions are satisfied, it is expected 
that our measurement scheme can be generalized to the 
multi-dimensional case. 
However, in many situations any physically available observable 
does not contain an eigenvector that is identical to a desired 
target state, and then our measurement scheme cannot be used 
for preparing that target state. 
Rather, as discussed in \cite{AshhabPRA2010,WisemanNature2011}, 
in such a case, a certain weak measurement may work for generating 
a target state. 
But a critical requirement for this measurement is that the 
measurement strength should be weakened when the state approaches 
to the target; 
then, the system can become fragile against unknown disturbance, 
as seen in Sec.~IV. 
Exploring an adaptive measurement method that overcomes this 
issue is an interesting future work.

We thank Yuzuru Kato for his contribution to Remark~(ii) in Sec.~II.


\end{document}